\documentstyle[aps,prl,twocolumn,epsf,floats,12pt]{revtex}

\draft
\begin{document}
\title{Composite Fermion Picture for Multi-Component Plasmas\\
       in 2D Electron-Hole Systems in a Strong Magnetic Field}
\author{
   \underline{A. W\'ojs}$^{1,2}$, 
   I. Szlufarska$^{1,2}$, 
   K.-S. Yi$^{1,3}$,
   P. Hawrylak$^4$,
   J. J. Quinn$^1$}
\address{\footnotesize\sl
   $^1$Department of Physics, 
       University of Tennessee, Knoxville, Tennessee 37996, USA \\
   $^2$Institute of Physics, 
       Wroclaw University of Technology, Wroclaw 50-370, Poland \\
   $^3$Physics Department, 
       Pusan National University, Pusan 609-735, Korea \\
   $^4$Institute for Microstructural Sciences, 
       National Research Council, Ottawa, Canada K1A 0R6\\[1em]}
\address{
   \footnotesize\rm\parbox{6.5in}{
   Low lying states of a 2D electron-hole system contain electrons 
   and one or more types of charged excitonic complexes.
   Binding energies and angular momenta of these excitonic ions,
   and the pseudopotentials describing their interactions with electrons
   and with one another are obtained from numerical studies of small
   systems.
   Incompressible fluid ground states of such multi-component plasmas
   are found in exact numerical diagonalizations.
   A generalized composite Fermion (CF) picture involving Chern--Simons
   charges and fluxes of different types is proposed and shown to predict
   the low lying states at any value of the magnetic field.
   }\\[-3ex]}
\maketitle
\vspace*{-6pt}
\paragraph*{Introduction.}
In a 2D electron-hole system in a strong magnetic field, the only bound 
complexes are neutral excitons $X^0$ and spin-polarized charged excitonic 
ions $X_k^-$ ($k$ excitons bound to an electron)
\cite{shields,wojs1,palacios,wojs2}.
Other complexes found at lower fields\cite{kheng} unbind due to a hidden 
symmetry\cite{lerner}.
The $X_k^-$ ions are long lived Fermions whose energy spectra contain 
Landau level structure\cite{wojs1,palacios,wojs2}.
By numerical diagonalization of small systems we can determine binding 
energies and angular momenta of the excitonic ions, and pseudopotentials 
which describe their interactions with electrons and with one another
\cite{wojs2}.
We show that a gas of $X_k^-$'s can form Laughlin\cite{laughlin}
incompressible fluid states\cite{wojs2}, but only for filling factors
$\nu_k\le(2k+1)^{-1}$ (in the following, subscript $k$ denotes $X_k^-$).
Multi-component plasmas containing electrons and $X_k^-$ ions of one 
or more different types can also form incompressible fluid states.
A generalized composite Fermion (CF) picture\cite{jain} is proposed 
to describe such a plasma\cite{wojs0}.
It requires the introduction of Chern--Simons\cite{lopez} charges and
fluxes of different types (colors) in order to mimic generalized 
Laughlin type correlations\cite{halperin}.
The predictions of this CF picture agree well with numerical results 
for systems containing up to eighteen particles.

\begin{figure}[t]
\epsfxsize=3.1in
\epsffile{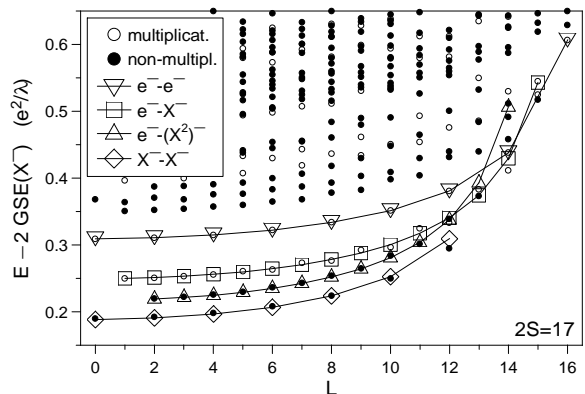}
\caption{ 
   Energy spectrum of the four electron and two hole system 
   at $2S=17$.}
\label{fig1}
\end{figure}

\paragraph*{Four Electron--Two Hole System.}
Understanding of the energy spectrum of this simple system is essential 
for our considerations.
Result of the numerical diagonalization in Haldane spherical geometry
\cite{haldane}, for the magnetic monopole strength $2S=17$, is shown 
in Fig.~\ref{fig1}.
Open and solid circles mark multiplicative and non-multiplicative 
states\cite{lerner}, respectively.
For $L<12$ there are four low lying bands, which we have identified,
in order of increasing energy, as two $X^-$'s, an electron and an 
$X_2^-$, an electron and an $X^-$ and a decoupled $X^0$, and finally 
two electrons and two decoupled $X^0$'s.
We find that the $X_k^-$ has an angular momentum $l_k=S-k$ in 
contrast to an electron which has $l_0=S$.
All relevant binding energies and pseudopotentials are also determined.
An important observation is that the pseudopotential of composite 
particles ($k>0$) is effectively infinite (hard core) if $L$ 
exceeds a particular value.
This is due to unbinding of ions at too small separation.
Once the maximum allowed $L$'s for all pairings are established, the four 
bands in Fig.~\ref{fig1} can be approximated by the pseudopotentials of 
electrons (point charges) with angular momenta $l_A$ and $l_B$, shifted 
by the appropriate binding energies (large symbols).

\paragraph*{Larger Systems}
We know from exact calculations for up to eleven electrons\cite{wojs3} 
that the CF picture correctly predicts the low lying states of the 
fractional quantum Hall systems.
The reason for this success is\cite{wojs3} the ability of the electrons 
in states of low $L$ to avoid large fractional parentage (FP)\cite{wojs3} 
from pair states associated with large values of the Coulomb pseudopotential.
In particular, for the Laughlin $\nu_0=1/3$ state, the FP from pair states 
with maximum pair angular momentum $L=2l_0-1$ vanishes.
We hypothesize that the same effect should occur for an $X^-$ system 
when $l_0=S$ is replaced by $l_1=S-1$.
We define an effective $X^-$ filling factor as $\nu_1(N,S)=\nu_0(N,S-1)$
and expect the incompressible $X^-$ states at all Laughlin and Jain 
fractions for $\nu_1\le1/3$.
States with $\nu_1>1/3$ cannot be constructed because they would have 
some FP from pair states forbidden by the hard core repulsion\cite{wojs2}.

\begin{figure}[t]
\epsfxsize=3.1in
\epsffile{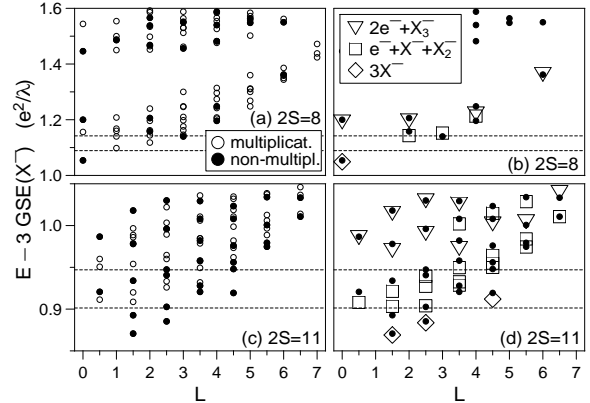}
\caption{
   Energy spectra of the six electron and three hole system 
   at $2S=8$ and 11.}
\label{fig2}
\end{figure}

Fig.~\ref{fig2} shows energy spectra of the $6e+3h$ system at $2S=8$ and 11.
Both multiplicative (open circles) and non-multiplicative (solid circles)
states are shown in frames (a) and (c).
In frames (b) and (d) only the non-multiplicative states are plotted,
together with the approximate spectra (large symbols) obtained by 
diagonalizing the system of three ions with the actual pseudopotentials 
appropriate to the three possible charge configurations: $3X^-$ (diamonds),  
$e^-+X^-+X_2^-$ (squares), and $2e^-+X_3^-$ (triangles).

Good agreement between the exact and approximate spectra in 
Figs.~\ref{fig2}b and \ref{fig2}d allows identification of the three 
ion states and confirms our conjecture about incompressible states of 
a $X^-$ gas.
States corresponding to different charge configurations form bands
At low $L$, the bands are separated by gaps, predominantly due to 
different total binding energies of different configurations.
The lowest state in each band corresponds to the three ions moving as 
far from each other as possible.
If the ion--ion repulsion energies were equal for all configurations 
(a good approximation for dilute systems), the two higher bands would 
lie above dashed lines, marking the ground state energy plus the 
appropriate difference in binding energies.
The low lying multiplicative states can also be identified as
$3e^-+3X^0$, $2e^-+X^-+2X^0$, $2e^-+X_2^-+X^0$, and $e^-+2X^-+X^0$.
The bands of three ion states are separated by a rather large gap from 
all other states, which involve excitation and breakup of composite 
particles.

\begin{figure}[t]
\epsfxsize=3.1in
\epsffile{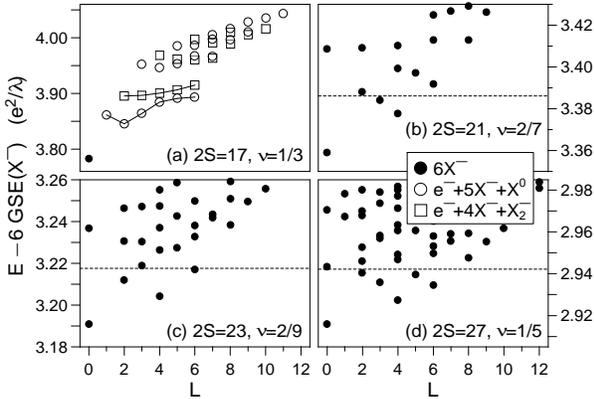}
\caption{
   Low energy spectra of different charge configurations of the 
   twelve electron and six hole system at $2S=17$, 21, 23, and 27.}
\label{fig3}
\end{figure}

The largest systems for which we performed exact calculations are the 
$6e+3h$ and $8e+4h$ systems at $2S$ up to 12 (Laughlin $\nu_1=1/5$ state 
of three $X^-$'s and one quasi-$X^-$-hole in the $\nu_1=1/3$ state of 
four $X^-$'s).
In each case the CF picture applied to the $X^-$ particles works well.
For larger systems the exact diagonalization becomes difficult.
For example, for the $12e+6h$ system we expect the $\nu_1=1/3$, 2/7, 2/9, 
and 1/5 incompressible states to occur at $2S=17$, 21, 23, and 27, 
respectively.
We managed to extrapolate the involved pseudopotentials making use of 
their regular dependence on $2S$, and use them, together with the binding 
energies, to determine approximate low lying bands in the energy spectra, 
as shown in Fig.~\ref{fig3}.
At $2S=17$, the only state of the $6X^-$ configuration is the $L=0$ 
ground state (filled circle); other $6X^-$ states are forbidden by the 
hard core.
The low lying states of other low energy configurations, $e^-+5X^-+X^0$ 
(open circles) and $e^-+4X^-+X_2^-$ (open squares) are separated from 
the $6X^-$ ground state by a gap.
At $2S=21$, 23, and 27, all features predicted by the CF picture occur 
for the $6X^-$ states.

\paragraph*{Generalized Composite Fermion Picture}
In order to understand all of the numerical results presented in 
Fig.~\ref{fig3}a, we introduce a generalized CF picture by attaching 
to each particle fictitious flux tubes carrying an integral number of 
flux quanta $\phi_0$.
In the multi-component system, each $a$-particle carries flux $(m_{aa}-1)
\phi_0$ that couples only to charges on all other $a$-particles and 
fluxes $m_{ab}\phi_0$ that couple only to charges on all $b$-particles,
where $a$ and $b$ are any of the types of Fermions.
The effective monopole strength\cite{jain,wojs3} seen by a CF of type 
$a$ (CF-$a$) is $2S_a^*=2S-\sum_b(m_{ab}-\delta_{ab})(N_b-\delta_{ab})$.
For different multi-component systems we expect generalized Laughlin 
incompressible states when all the hard cores are avoided and CF's of 
each type fill completely an integral number of their CF shells.
In other cases, the low lying multiplets will contain different types
of quasiparticles (QP-$a$, QP-$b$, \dots) or quasiholes (QH-$a$, QH-$b$, 
\dots) in the neighboring incompressible state.

Our multi-component CF picture can be applied to the system of excitonic 
ions, where the CF angular momenta are given by $l_k^*=|S_k^*|-k$.
As an example, let us consider Fig.~\ref{fig3}a and make the following 
CF predictions.
For six $X^-$'s we obtain the Laughlin $\nu_1=1/3$ state at $L=0$.
Because of the $X^-$-$X^-$ hard core, it is the only state of this 
configuration.
For the $e^-+5X^-+X^0$ configuration we set $m_{11}=3$ and $m_{01}=1$, 
2, and 3.
For $m_{01}=1$ we obtain $L=1$, 2, $3^2$, $4^2$, $5^3$, $6^3$, $7^3$, 
$8^2$, $9^2$, 10, and 11;
for $m_{01}=2$ we obtain $L=1$, 2, 3, 4, 5, and 6; and
for $m_{01}=3$ we obtain $L=1$.
For the $e^-+4X^-+X_2^-$ configuration we set $m_{11}=3$, $m_{02}=1$, 
$m_{12}=3$, and $m_{01}=1$, 2, or 3.
For $m_{01}=1$ we obtain $L=2$, 3, $4^2$, $5^2$, $6^3$, $7^2$, $8^2$, 9, 
and 10;
for $m_{01}=2$ we obtain $L=2$, 3, 4, 5, and 6; and 
for $m_{01}=3$ we obtain $L=2$.
Note that the sets of multiplets obtained for higher values of 
$m_{01}$ are subsets of the sets obtained for lower values; 
we would expect them to form lower energy bands since they avoid 
additional large values of $e^-$-$X^-$ pseudopotential.
As marked with lines in Fig.~\ref{fig3}a, this is indeed true for the 
states predicted for $m_{01}=2$.
However, the states predicted for $m_{01}=3$ do not form separate bands.
This is because $e^-$-$X^-$ pseudopotential increases more slowly than 
linearly as a function of $L(L+1)$ in the vicinity of $L=l_0+l_1-m_{01}$;
in such case the CF picture fails\cite{wojs3}.

The agreement of our CF predictions with the exact spectra of different
systems, as in Figs.~\ref{fig2} and \ref{fig3}, is really quite remarkable 
and strongly indicates that our multi-component CF picture is correct.
We are actually able to confirm predicted Laughlin type correlations
\cite{halperin} in the low lying states by calculating their FP 
coefficients\cite{wojs3}.
In view of the results obtained for many different systems that we were 
able to treat numerically, we conclude that if exponents $m_{ab}$ are 
chosen correctly, the CF picture works well in all cases.

\paragraph*{Summary}
Low lying states of electron-hole systems in a strong magnetic field
contain charged excitonic ions $X_k^-$ interacting with one another
and with electrons.
For different combinations of ions occurring at low energy, we introduced 
general Laughlin type correlations into the wavefunctions and demonstrated 
formation of incompressible fluid states of such multi-component plasmas 
at particular values of the magnetic field.
We also proposed a generalized multi-component CF picture and successfully 
predicted lowest bands of multiplets for various charge configurations 
at any value of the magnetic field.
It is noteworthy that the fictitious Chern--Simons fluxes and charges 
of different types or colors are needed in the generalized CF model.
This strongly suggests that the effective magnetic field seen by the 
CF's does not physically exist and that the CF picture should be 
regarded as a mathematical convenience rather than physical reality.
Our model also suggests an explanation of some perplexing observations 
found in photoluminescence, but this topic will be addressed in a separate 
publication.

We thank M. Potemski for helpful discussions.
AW and JJQ acknowledge partial support from the Materials Research 
Program of Basic Energy Sciences, US Department of Energy.
KSY acknowledges support from Korea Research Foundation.

\end{document}